\documentclass{iopart}
\usepackage{graphicx}
\usepackage[dvips]{color}

\begin{document}

\title{Displacement fields of point defects in two-dimensional colloidal crystals}

\author{Wolfgang Lechner, Elisabeth Sch{\"o}ll-Paschinger, and Christoph Dellago}
\relax

\address{Faculty of Physics, University of Vienna, Boltzmanngasse 5, 1090 Vienna, Austria}

\date{\today}

\begin{abstract}
Point defects such as interstitials, vacancies, and impurities in otherwise perfect crystals induce complex displacement fields that are of long-range nature. In the present paper we study numerically the response of a two-dimensional colloidal crystal on a triangular lattice to the introduction of an interstitial particle. While far from the defect position the resulting displacement field is accurately described by linear elasticity theory, lattice effects dominate in the vicinity of the defect. In comparing the results of particle based simulations with continuum theory, it is crucial to employ corresponding boundary conditions in both cases. For the periodic boundary condition used here, the equations of elasticity theory can be solved in a consistent way with the technique of Ewald summation familiar from the electrostatics of periodically replicated systems of charges and dipoles. Very good agreement of the displacement fields calculated in this way with those determined in particle simulations is observed for distances of more than about 10 lattice constants. Closer to the interstitial, strongly anisotropic displacement fields with exponential behavior can occur for certain defect configurations. Here we rationalize this behavior with a simple bead-spring that relates the exponential decay constant to the elastic constants of the crystal.  
\end{abstract}

\maketitle


\section{Introduction}

The properties of crystalline substances often crucially depend on the structure and dynamics of imperfections of the crystal lattice. In particular, point defects such as interstitials and vacancies play a pivotal role in determining the stability, transport properties, growth characteristics, and mechanical behavior of materials. Recent impressive experimental advances, such as optical tweezers and confocal microscopy \cite{SCIENCE_GASSER,CONFOCAL}, now permit to study the fundamental properties of point defects in condensed matter systems with ``atomistic'' space and time resolution. 

Recently, a number of experimental studies have focused on the structure and dynamics of point defects in two-dimensional assemblies of micrometer sized colloidal particles \cite{PERTSINIDIS_NATURE,PERTSINIDIS_NJP,PERTSINIDIS_PRL} and, in particular, on their effective interactions \cite{MELTING_2D,MARET,GRUENBERG_PAIR}. In studying such defect interactions the question arises to which degree they can be rationalized in terms of continuum elastic theory. As a first step towards answering this question, in this article we investigate numerically the disturbances caused by isolated interstitial particles and compare the results with the predictions of continuum theory. In carrying out such a comparison, it proves crucial that in solving the equations of elasticity theory boundary conditions are used that match those of the simulations. For the periodic boundary conditions usually applied in computer simulations, the displacement fields of single defects can be determined using the technique of Ewald summation familiar from electrostatics \cite{LEEUW,NEUMANN}. While elasticity theory properly describes the disturbances and interactions created by lattice imperfections on a larger scale, discrete lattice effects dominate on spatial scales of the order of few lattice constants. 

The remainder of this paper is organizes as follows. In Sec. \ref{sec:model} we define the model and describe the numerical methods.  The treatment of point defects in a two-dimensional elastic continuum is discussed in Sec. \ref{sec:elasticity} and comparison with the numerical results is discussed in Sec. \ref{sec:results}. For certain defect configurations one observes an exponential rather than algebraic decay of the displacement fields. This behavior can be understood in terms of a simple bead-spring model introduced in Sec. \ref{sec:spring} with parameters related to the elastic constants of the material. Some concluding remarks are provided in Sec. \ref{sec:discussion}. 
 

\section{Simulations}
\label{sec:model}  

In this paper we study a two-dimensional crystal of soft particle interacting via the Gaussian potential \cite{STILLINGER_GCM,GAUSSIAN_CORE_PHASE,GAUSSIAN_SOFTLY_PHASE}
\begin{equation}
v(r) = \epsilon \exp(-r^2/\sigma^2),
\end{equation}
where $r$ is the inter-particle distance and $\epsilon$ and $\sigma$ set the energy and length scales, respectively. In the following, energies are measured in units of $\epsilon$ and distances in units of $\sigma$. This so-called Gaussian core model, used here as a generic model for a system of soft spheres, is a realistic description for the short-ranged effective interactions between polymer coils in solution \cite{FLORY}. In three dimensions, the Gaussian core model can exist as a fluid, a bcc- and an fcc-solid depending on temperature and density \cite{GAUSSIAN_CORE_PHASE}. In two dimensions, the perfect triangular lattice is the lowest energy structure of Gaussian core particles at all densities \cite{GCM_STILL_2d_1}. Computer simulations indicate that also in this system of purely repulsive particles point defects such as interstitials, vacancies or impurity particles of different size display attractive (as well as repulsive) interactions both in two and three dimensions \cite{LD_PREP}.

To study the displacement field of a single interstitial numerically, we prepare a configuration of particles arranged on the sites of a perfect lattice configuration and insert an extra particle of the same species. After insertion, the system is relaxed to a new minimum energy configuration by steepest descent minimization, i.e., we study the defect structure at $T=0$. Typically, $70.000$ steepest descent steps are carried out. The system we study here consists of  $N=199.680$ Gaussian core particles (without the extra particle) at a number density of $\rho = 0.6 \sigma^{-2}$ corresponding to a lattice constant $a=(2/\sqrt{3} \rho)^{1/2}= 1.3872 \sigma$. Periodic boundary conditions apply to the simulation box of length $L_x = 416 a$ and height $L_y = (\sqrt{3}/2) 480\, a=415.692 a$. The aspect ratio of the almost square simulation box is $L_y/L_x=0.99926$. 

We quantify the perturbation caused by the defect in terms of the displacement field \cite{LANDAU_LIFSCHITZ}
\begin{equation}
\label{equ:displacementfield}
{\bf u}({\bf r}_i) \equiv {\bf r}'_i - {\bf r}_i.
\end{equation}
Here, ${\bf r}'_i$ and ${\bf r}_i$ denote the position of particle $i$ with and without the defect, respectively.  As we will see in the following sections, simple point defects generate remarkably intricate displacement patterns that can be understood in terms of elasticity theory only on large length scales.

At $T=0$, the elastic constants describing the macroscopic response of the system to perturbations can be calculated as a function of density from simple lattice sums. For a density of $\rho=0.6\sigma^2$, the Lam\'e coefficients (see Sec. \ref{sec:elasticity}) of the perfect triangular lattice have values $\lambda=1.1487\, \epsilon \sigma^{-2}$ and $\mu=0.06018 \,\epsilon \sigma^{-2}$. At this density, the pressure is $p=0.5442 \, \epsilon \sigma$ and the energy density is $e=0.2691 \,\epsilon \sigma^{-2}$ corresponding to an energy per particle of $E/N=0.4485 \epsilon$. The bulk modulus, which in two dimensions is related to the  the  Lam\'e coefficients by $K=\lambda + \mu$, has a value of $K=1.2089 \epsilon \sigma^{-2}$.


\section{Elasticity Theory}
\label{sec:elasticity}

While close to a point defect the displacement field is highly anisotropic and strongly dependent on the atomistic details of the interactions, for large distances elasticity theory is expected to be valid. The differential equations describing the equilibrium of an elastic continuum are usually expressed in terms of the strain tensor \cite{LANDAU_LIFSCHITZ}
\begin{equation}
\epsilon_{ij}({\bf r})  =  \frac{1}{2} \left(\frac{\partial u_i}{\partial r_j} + \frac{\partial u_j}{\partial r_i} \right),
 \end{equation}
where $u_i$ denotes the $i$-component of the displacement ${\bf u}$ and $r_i$ the $i$-th component of the position ${\bf r}$. For a given external volume force ${\bf f}({\bf r})$ with components $f_i$ acting on an isotropic system such as a crystal on a triangular lattice, Hook's law leads to the equilibrium condition for the strain: 
\begin{equation}
\lambda \frac{\partial}{\partial r_i}\epsilon_{kk} + 2\mu \frac{\partial \epsilon_{ij}}{\partial r_j}+f_i = 0.
\label{equ:equil_strain}
\end{equation}
Here, $\lambda$ and $\mu$ are the so-called Lam\'e coefficients and summation over repeated indices is implied. Solving this equation for a singular force yields the Green's function from which the response of the elastic continuum to an arbitrary force can be obtained by integration. 

To model the displacement field caused by the introduction of point defects using linear continuum elasticity theory, we determine the displacement field caused by two pairs of opposing forces, one pair acting along the $x$-axis and the other one along the $y$-axis \cite{ESHELBY,ESHELBY_ACTA,SCATTERGOOD}. This idealized model of a defect is equivalent to inserting a small circular inclusion into a hole of different size \cite{ESHELBY_ACTA}. Each force of this pair is of equal magnitude $F$ but with opposite sign acting on two points separated by a small distance $h$. Such a force pair exerts a zero net force on the material. In the limit $h \rightarrow 0$ where the force $F \rightarrow \infty$ in a way such that $Fh$ remains constant, the equilibrium condition for the displacement can be written as
\begin{equation}
(\lambda + 2\mu) \frac{\partial u_j}{\partial r_j}=Fh \delta ({\bf r}).
\end{equation}
Assuming that the displacement can be written as the derivative of a potential,
\begin{equation}
u_i=\frac{\partial \phi}{\partial r_i}
\label{equ:displacement_phi}
\end{equation}
one obtains 
\begin{equation}
\label{equ:Poisson}
\Delta \left(-\frac{\lambda + 2\mu}{Fh} \phi \right)=-\delta ({\bf r}).
\end{equation}
This equation is the Poisson equation of electrostatics with a singular disturbance. Since, as noted above,  $K({\rm r})=-\ln (r)/2 \pi$ is a solution of $\Delta K = -\delta ({\bf r})$ (see, for instance, Ref. \cite{COURANT}), we obtain the Green's function
\begin{equation}
\label{equ:phi_logarithmic}
\phi(r)=\frac{Fh}{2\pi(\lambda + 2\mu)} \ln (r)
\end{equation}
from which the displacement field ${\bf u}({\bf r})$ follows by differentiation according to Equ. (\ref{equ:displacement_phi}),
\begin{equation}
u_i=\frac{Fh}{2\pi(\lambda+2\mu)}\frac{r_i}{r^2}.
\end{equation}

In comparing the results of particle simulations with those of elasticity theory it is important to realize that the displacement fields predicted by continuum theory are of a long-range nature. Therefore, it is crucial that corresponding boundary conditions are used in both cases. All simulations discussed in this paper are done with periodic boundary conditions in order to minimize surface effects and preserve the translational invariance of the perfect lattice. Hence,  also the continuum calculations need to be carried out with periodic boundary conditions. 
            
Since the defect fields for the infinite material are long-ranged, the displacement field in the periodic system cannot be obtained by simply summing up the contributions of the periodic images. In fact, such a naive summation of the contribution of all image defects diverges. A more appropriate treatment that avoids this problem consists in determining the Green's function of the Poisson equation (\ref{equ:Poisson}) for periodic boundary conditions. In this case, the solution of this equation in two dimensions, known from electrostatics \cite{LEEUW,NEUMANN}, can be written as Ewald sum of a logarithmic potential embedded in a neutralizing background,
\begin{eqnarray}
	\label{equ:EwaldPotential}
	\phi({\bf r}) & = &\frac{Fh}{2\pi(\lambda+2\mu)}\left\{\frac{1}{2}\sum_{\bf l}E_i[-\eta^2 |{\bf r}+{\bf l}|^2] \right.
	\nonumber\\
	 & & 	\hspace{1.5cm} \left. - \frac{2 \pi}{A} \sum_{{\bf k} \neq 0} \frac{e^{-k^2/4\eta^2}}{k^2} \cos({\bf k}\cdot {\bf r})  +\frac{\pi}{2 \eta^2 A} \right\}.
\end{eqnarray}
Here,  $E_i(x)=\int_{-\infty}^x (e^t /t) \, dt$ is the exponential integral. The first sum is over all lattice vectors ${\bf l}$ in real space and the second sum is over all reciprocal vectors ${\bf k}$ in Fourier space. The adjustable parameter $\eta$, set to a value of $\eta=6/L_x$ here, determines the rate of convergence of the two sums and $A$ is the area of the rectangular simulation cell. The Fourier space sum can be evaluated accurately using about 2,500 reciprocal space vectors. From Equ. (\ref{equ:EwaldPotential}) for the scalar function $\phi({\bf r})$ the displacement field of a point defect in a system with periodic boundary conditions is found by differentiation,
\begin{eqnarray}
	\label{equ:EwaldDisplacement}
	u_i({\bf r}) & = & \frac{Fh}{2\pi (\lambda + 2\mu)}\left\{
	\sum_{\bf l} e^{-\eta^2 |{\bf r}+{\bf l}|^2} \frac{r_i+l_i}{|{\bf r}+{\bf l}|^2} \right. \nonumber \\
 &  &	\hspace{1.5cm}+\left. 
	\frac{2\pi}{A} \sum_{{\bf k} \neq 0} \frac{e^{-k^2/4\eta^2}}{k^2} k_i
 \sin(\mathbf{k}\cdot \mathbf{r})\right\}.
\end{eqnarray}
For the systems considered in this paper, the real space sum may be truncated after the first term. Since the value of $Fh/2\pi (\lambda + 2\mu)$ is undetermined,  the parameter $\gamma \equiv Fh / 2\pi (\lambda + 2\mu)$ is treated as a fit parameter in the following. The Ewald sums of the above equations describe the effects of ``image defects'' at the center of the periodically replicated domains. 


\section{Results}
\label{sec:results}
First, we study the displacement field of a single interstitial. To generate such a defect, we insert an extra particle of the same species into a perfect 2d-crystal on a triangular lattice. After insertion, the system is relaxed to a new minimum energy configuration by steepest descent minimization, i.e., we study the defect structure at $T=0$. Typically, $70.000$ steepest descent steps are carried out. In each step each particle is moved in the direction of the force acting on the particle where the absolute value of the displacement in chosen to be small enough to ensure that the energy of the system decreases in each step. 

\begin{figure}[h]
  \begin{minipage}[t][7.35cm][t]{0.44cm}
    (a)
  \end{minipage}
  \begin{minipage}[t][7.35cm][b]{3.78cm}
    \includegraphics[width=3.78cm]{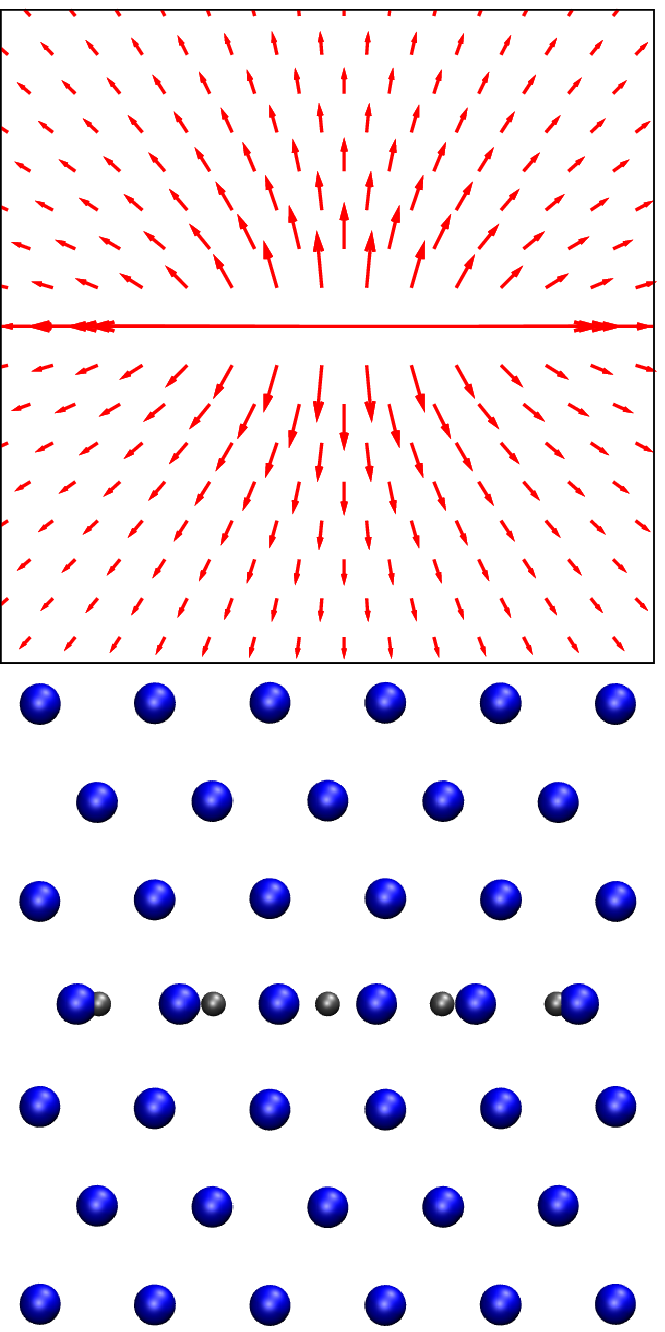} 
  \end{minipage}
  \begin{minipage}[t][7.35cm][t]{0.46cm}
    (b)
  \end{minipage}
  \begin{minipage}[t][7.35cm][b]{3.78cm}
    \includegraphics[width=3.78cm]{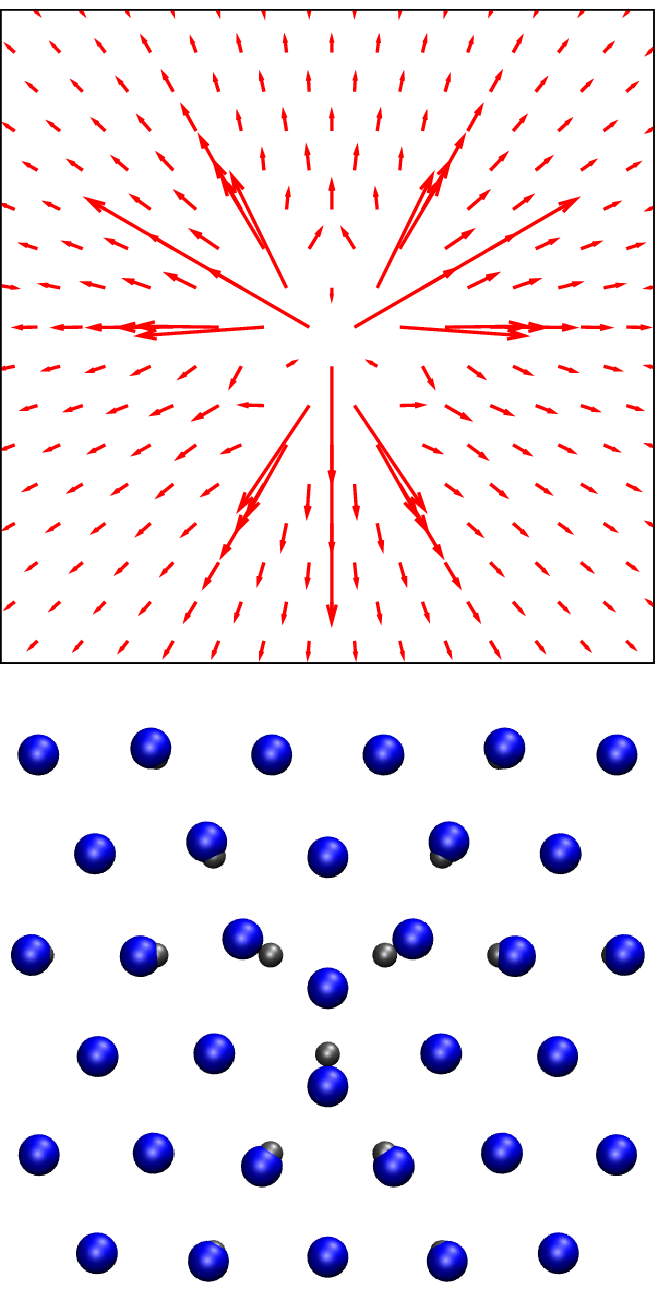} 
  \end{minipage}
  \begin{minipage}[t][7.35cm][t]{0.44cm}
    (c)
  \end{minipage}
  \begin{minipage}[t][7.35cm][b]{3.78cm}
    \includegraphics[width=3.78cm]{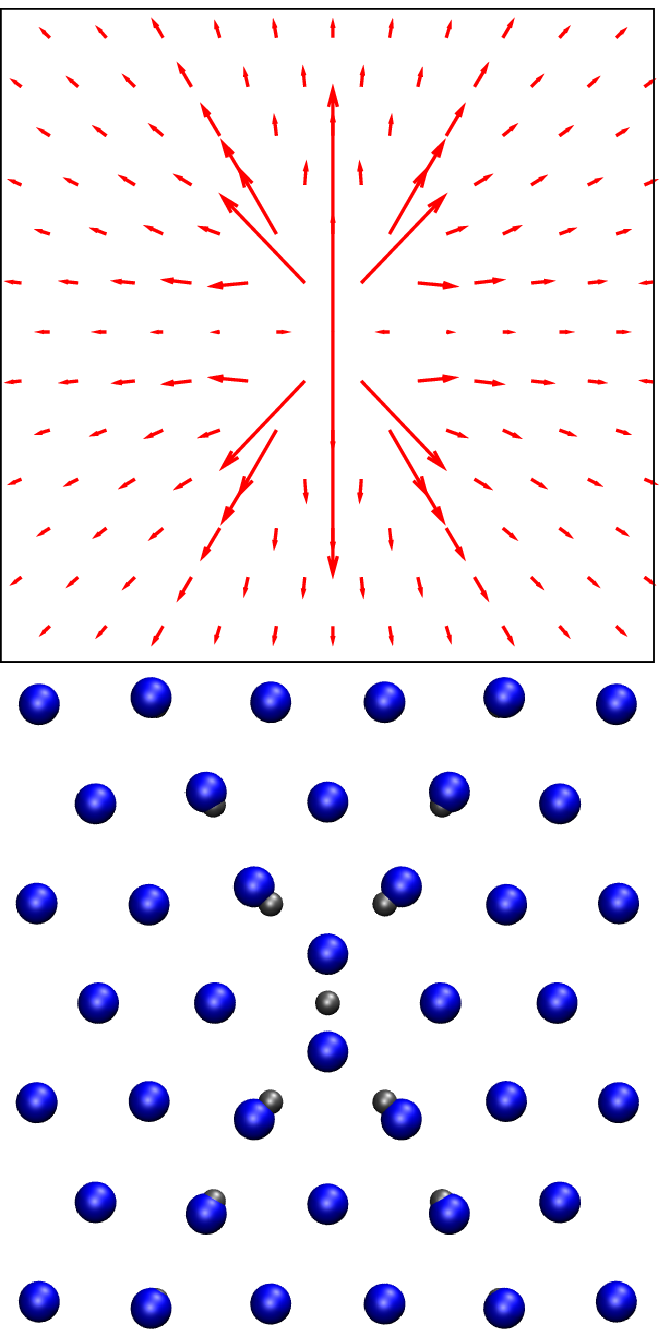} 
  \end{minipage}
  \caption{Displacement fields (top) and local defect configurations (bottom) for the $I_2$ (a), the $I_3$ (b) and the $I_d$ (c) configurations. The length of the arrows representing the displacements of the particles from their position in the perfect lattice are exaggerated for better visibility. In the figures at the bottom the blue spheres represent the particles and the small gray spheres indicate the position of the lattice sites of the perfect crystal.}
  \label{fig:displacementinter}
\end{figure}

The extra particle can deform the crystal in different ways \cite{PERTSINIDIS_NJP} and produces displacement fields of different symmetries (see Fig. \ref{fig:displacementinter}). In one configuration, called $I_2$ interstitial or crowdion and shown in Fig. \ref{fig:displacementinter}a, the additional particle pushes one particular particle of the crystal out of its equilibrium position. Both the original particle and the additional particle arrange themselves at equal distance around the lattice position of the original particle. The displacement pattern arising for this type of interstitial has two-fold symmetry and, of course, occurs in all three low-index lattice directions with equal probability. One may suspect that this defect configuration, with a symmetry that differs from the symmetry of the underlying triangular lattice, is caused by the rectangular periodic boundary conditions that are applied to the system. To rule out this possibility, we have repeated the calculation with hexagonal periodic boundary conditions obtaining the same result. 

Another low-energy defect configuration is the $I_3$ interstitial with three-fold symmetry (see Fig. \ref{fig:displacementinter}b). In this case the interstitial particle is located at the center of a basic lattice triangle and pushes its neighbors outward from their original positions. A third important interstitial configuration is the $I_d$ interstitial or dumbbell interstitial shown in Fig. \ref{fig:displacementinter}c. In the 2d Gaussian-core model under the conditions studied here the $I_2$ pattern has a slightly lower energy than the $I_3$ interstitial and the $I_d$ interstitial. The energy difference between a $I_2$ and a $I_3$ interstitial is $0.000674 \epsilon$ and the difference between $I_2$ and $I_d$ is $0.000665 \epsilon$. All three displacement patterns are important for the diffusion of interstitials. An $I_2$ interstitial is very mobile in the direction of its main axis. The $I_3$ and $I_d$ forms are visited as intermediate configurations when the $I_2$ interstitial changes the orientation of its main axis and hence its direction of motion \cite{REICHHARDT}. 

Next, we compare the displacement fields determined numerically with the predictions of elasticity theory. In particular, we verify to which extent the $1/r$-behavior modulated by the periodic boundary conditions and embodied in Equ. (\ref{equ:EwaldDisplacement}) is realized in the particle system. The complex displacement patterns  of the various interstitial configurations shown in Fig. \ref{fig:displacementinter} obviously differ from this expectation, at least near the defect, and indicate that continuum theory is not applicable in this region. Far away from the defect, however, the perturbation caused by the defect is small and the response of the material should be described accurately by linear elasticity theory.   

\begin{figure}[h,floatfix]
\centerline{\includegraphics[width=8.0cm]{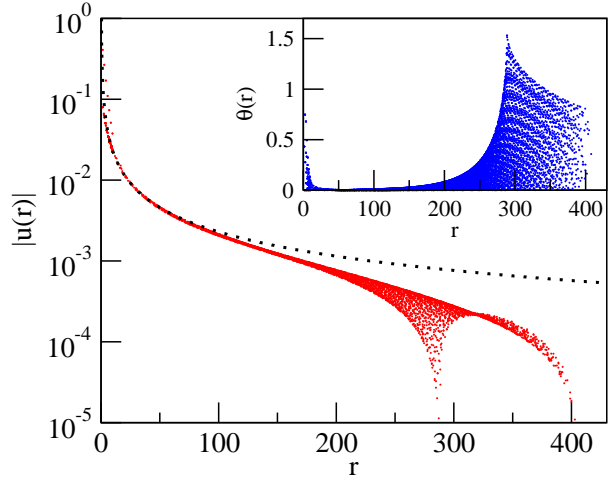}}
\caption{Displacement magnitude $|{\bf u}({\bf r})|$ as a function of distance from the defect $r$ for the $I_2$ interstitial. Each red dot corresponds to one particle. The solid line represents the $\gamma/r$ behavior. Here, $\gamma = 0.2291\sigma^2$ was used as this value yields the best fit of the results obtained vie Ewald summation to the results of the particle simulations in the far field. Inset: angle $\theta$ between the displacement vector ${\bf u}$ the position vectors ${\bf r}$ as a function of the distance $r$ from the defect site.
}
\label{fig:disp_theta}
\end{figure}

The magnitude $|{\bf u}({\bf r})|$ of the displacement vector ${\bf u}({\bf r})$ is shown as a function of the distance from the interstitial in Fig. \ref{fig:disp_theta} for the $I_2$ defect configuration. Each point in the figure corresponds to one individual particle. For short distances, the displacement magnitude is not a unique function of the distance $r$ reflecting the anisotropic nature of the defect. For larger distances, however, the displacement magnitude is mostly determined by the distance $r$. Eventually, however, the periodic boundary conditions lead to a spread of the displacement magnitude for even larger distances and a splitting into two branches corresponding to the $x$- and $y$-directions and the directions along the diagonals, respectively. In the regime where ${\bf u}({\bf r})$ behaves isotropically, the displacement follows the approximately $1/r$-form predicted by elasticity theory for a point defect in an infinitely extended medium. The orientation of the displacement vector ${\bf u}({\bf r})$, depicted in the inset of Fig. \ref{fig:disp_theta}, behaves in an analogous way. The angle $\theta$ between ${\bf u}({\bf r})$ and the position vector ${\bf r}$, shown as a function of the distance $r$ from the defect, is not a unique function of $r$ near the defect. For larger $r$, $\theta$ vanishes indicating that in this distance regime the displacement vector points straight away from the defect. At even larger distances, the periodic boundary conditions imposed on the system eventually cause the angle $\theta$ to spread again.

\begin{figure}[h,floatfix]
\centerline{\includegraphics[width=8.0cm]{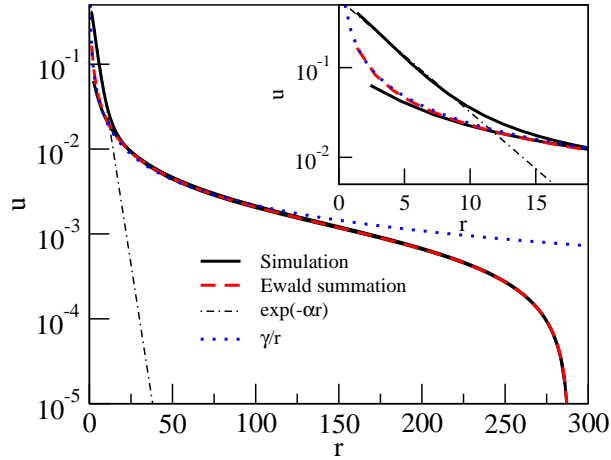}}
\caption{Displacement components $u_x$ and $u_y$ of the $I_{2}$ interstitial as s function the of the distance along the $x$-axis and $y$-axis, respectively (solid lines). Here, the direction of largest displacement of the $I_{2}$ defect is oriented in $x$-direction. Also plotted is the displacement computed from continuum theory according to Equ. (\ref{equ:EwaldDisplacement}) (dashed line), simple $1/r$-behavior (dotted line) and the displacement obtained for the simple mechanical model described in the main text (dash-dotted line). The inset shows the region close to the defect location. As in Fig. \ref{fig:disp_theta} a defect strength of $\gamma = 0.2291 \sigma^2$ was used for the evaluation of the displacement from elasticity theory. }
\label{fig:absdisplacement}
\end{figure}

The displacement fields calculated according to Equ. (\ref{equ:EwaldDisplacement}) and numerically for an interstitial in the $I_2$ configuration are compared in Fig.\ref{fig:absdisplacement}. In this figure, the displacement components $u_x$ and $u_y$ are  depicted as a function of the distance from the defect along the $x$-axis and $y$-axis, respectively. The prediction of continuum theory, calculated using the Ewald summation of Equ. (\ref{equ:EwaldDisplacement}), agrees well with the displacement field of the particle system for distances larger than about 10 lattice constants. 


\section{Harmonic Model}
\label{sec:spring}

Near the defect, the predictions of continuum theory differ from the simulation results. The deviation is particularly pronounced in the direction of the main axis of distortion of the $I_2$ defect, in which the displacement appears to decay exponentially up to a distance of about $\approx 10 \,a$. This unexpected exponential behavior can be understood in terms of a simple model with harmonic interactions. This model consists of a one-dimensional chain of particles in which each particle is connected to its two neighbors with springs of force constant $k_1$ (except the first and last particle, which are coupled only to their neighbors on the right and left, respectively). In addition, each particle is attached to a fixed lattice position with another spring of force constant  $k_2$. The Hamiltonian of this system is
\begin{equation}
\label{equ:simplemodelhamiltonian}
\mathcal{H} = \frac{k_1}{2} \sum_{j=0}^{N} (x_{j+1} - x_j - b)^2 + \frac{k_2}{2} \sum_{j=0}^{N+1} (x_j - b j)^2,
\end{equation}
where $N+2$ is the number of particles, $x_j$ is the position of particle $j$ and $b$ is the equilibrium distance between two neighboring particles. In the minimum energy configuration of this chain, the particles are arranged such that $x_j = jb$. We now imagine that particle $0$ is pushed to the right by a distance of $u_0$ while particle $N+1$ is kept fixed at $x_{N+1}=(N+1)b$. If the system is then relaxed to a new energy minimum, all other particles will be displaced from their original positions too. For this simple model, the response of the system to the displacement of the first particle can be calculated analytically by direct matrix inversion (see \ref{ap:springmodel}). In the large $N$ limit, one finds that the displacement of the particles from their original position decays exponentially with their position,
\begin{equation}
\label{equ:displacementmechanical}
u_j = u_0 \exp(-\alpha j),
\end{equation}
where $u_j$ is the displacement of particle $j$ due to the forced displacement $u_0$ of the first particle. The decay constant $\alpha$ is related to the force constants of the model by 
\begin{equation}
\label{equ:lambdamechanical}
\alpha = \cosh^{-1}\left(1 + \frac{k_2}{2k_1} \right).
\end{equation}

To compare the prediction of this simple model with the simulation results we have to determine the force constants $k_1$ and $k_2$ felt by the particles in the main axis of the defect. While the force constant $k_1$ arises from interactions within this main axis, the force constant $k_2$ is related to interactions of the particles in the main axis with those from adjacent rows. Accordingly, we determine $k_1$ by calculating numerically the energy change caused by slightly displacing one single particle in a one-dimensional row of otherwise fixed Gaussian core particles without the presence of the neighboring rows. The distance of the particles in the row is chosen to be equal to the lattice constant at the density $\rho=0.6 \sigma^{-2}$ considered throughout the paper. From  the energy as a function of the displacement one obtains a force constant of $k_1 = 0.015 \epsilon / \sigma^2$. To determine the force constant $k_2$ we calculate the energy change caused by translating a whole row of particles in the perfect crystal. The particles in the row are fixed with respect to each other and the remaining particles are kept at their lattice positions. From the energy change per moved particle a force constant of  $k_2 = 0.0013 \epsilon / \sigma^2$ follows.  The decay constant of $\alpha = 0.29$ calculated according to Equ. (\ref{equ:lambdamechanical}) with these force constants is in perfect agreement with the computer simulation results shown in Fig. \ref{fig:absdisplacement}. 

For a system in which only nearest neighbor interactions are important, the force constants $k_1$ and $k_2$ can be simply related to the bulk modulus $K$ and the shear modulus $\mu$. Then, the force constant $k_1$ is given by 
\begin{equation}
k_1=2v''(a),
\end{equation}
where $v(a)$ is the pair potential at distance $a$. Since in this case the elastic moduli are given by
\begin{equation}
K=\frac{\sqrt{3}}{2}\left\{v''(a)-\frac{v(a)}{a}\right\}
\end{equation}
and
\begin{equation}
\mu=\frac{\sqrt{3}}{4}\left\{v''(a)+3\frac{v(a)}{a}\right\}
\end{equation}
one obtains 
\begin{equation}
k_1=\frac{2}{\sqrt{3}}\left( \mu + \frac{3K}{2}\right).
\end{equation}
To the extent that the response of the system to shear is determined by the interaction of neighboring parallel rows of particles, the energy density caused by shifting a whole row of atoms between two fixed ones is the same as that of a shear of appropriate magnitude. Accordingly, the force constant $k_2$ is related to the shear modulus by 
\begin{equation}
k_2 = \frac{4}{\sqrt{3}}\mu.
\end{equation}
This expression remains also valid if interactions beyond nearest neighbors are included between adjacent rows of particles. In terms of the elastic constants, the constant $\alpha$ describing the exponential decay of the displacement field along the principal axis can be expressed as 
\begin{equation}
\alpha = \cosh^{-1}\left(1 + \frac{2\mu}{2\mu+3K} \right),
\end{equation}
or, in terms of the Poisson ratio $\nu$,
\begin{equation}
\alpha = \cosh^{-1}\left(\frac{7-\nu}{5+\nu} \right).
\end{equation}
For a density of $\rho=0.6 \sigma^{-2}$, inserting the Poisson ratio of $\nu = 0.905151$ determined from a simple lattice sum yields $\alpha\approx 0.25$, only slightly different from the correct value $\alpha \approx 0.29$. This deviation occurs, because in the Gaussian core model at the density  $\rho=0.6 \sigma^{-2}$ interactions between non-nearest neighbor particles are important in determining the elastic constants (in fact, considering only nearest neighbors would produce a negative shearing modulus $\mu$ in this case). For systems, in which only nearest neighbor interactions are relevant, the above expression is expected to hold accurately.


\section{Conclusion}
\label{sec:discussion}

Point defects in two-dimensional crystals, such as interstitials and vacancies, can assume configurations with symmetries that vary from the symmetry of the underlying triangular lattice. While close to the defect the displacement field is highly anisotropic and strongly dependent on the atomistic details of the interactions, for large distances elasticity theory, which predicts isotropic behavior, is valid. For the particular I$_2$ interstitial configuration, the displacement decreases exponentially with distance along the main defect axis. The decay constant is simply related to the material properties via the Poisson ratio, which measures the ratio between transversal and axial strain upon stretching. In comparing the displacement fields computed from particles simulations with  those obtained with continuum elasticity theory it is crucial to use equivalent boundary conditions in both cases. Since particle simulations are usually carried out with periodic boundary conditions, also the differential equations of elasticity theory need to be solved for a periodic system. We have shown here that Ewald summation, a technique routinely used in computer simulations to determine the electrostatic interactions of charges and dipoles, can be used for this purpose. In this method the sum over all interactions with periodic image defects is split into two sums in real space and reciprocal space, respectively. This particular treatment of the long-ranged nature of displacement fields effectively introduces a neutralizing background that leads to convergent sums. Note that exactly the same expression apply also to a system that is enclosed in a rigid container. Outside a core region near the defect, displacement patterns determined using such Ewald summation agree perfectly with those calculated in particle simulations.   


\ack 
The authors would like to thank Christos Likos, Martin Neumann, and Andreas Tr\"oster and for useful discussions. This research was supported by the University of Vienna through the University Focus Research Area {\em Materials Science} (project ``Multi-scale Simulations of Materials Properties and Processes in Materials'').



\newpage

\begin{appendix}

\section{}
\label{ap:springmodel}

In this appendix we calculate the response at $T=0$ of the one-dimensional bead-spring model of Sec. \ref{sec:spring} to a forced displacement of the first particle in the chain. The potential energy of the $N+2$ particles, located at positions $x_j$, is given by
\begin{equation}
\label{equ:simplemodelhamiltonianA}
\mathcal{H}(x) = \frac{k_1}{2} \sum_{j=0}^{N} (x_{j+1} - x_j - b)^2 + \frac{k_2}{2} \sum_{j=0}^{N+1} (x_j - b j)^2,
\end{equation}
where $b$ is the equilibrium distance and $k_1$ and $k_2$ are force constants. The vector $x={x_0, x_1, \cdots, x_{N+1}}$ includes the positions of all particles. Minimizing the potential energy with respect to the particle positions $x_j$ by requiring that 
\begin{equation}
\label{equ:minpot}
\left.\frac{\partial {\mathcal H}(x)}{\partial x_j}\right|_{x=\overline x} = 0
\end{equation}
for all $j$, one finds that at the potential energy minimum the particle positions are $\overline x_j = b j$. We now displace particle $0$ by an amount $u_0$ from its original position $\overline x_0=0$ and keep particle $N+1$ fixed at position $(N+1)b$. If we hold particle $0$ at this new position while minimizing the potential energy, all particles from $1$ to $N$ will move to new equilibrium positions. Thus, the minimum energy configuration of the system is a function of the displacement $u_0$ of particle $0$, which may be viewed as a parameter that is controlled externally and perturbs the system. To make this distinction between the displacement of particle $0$ and that of all other particles more explicit, we denote $u_0$ with an extra symbol, $\xi=u_0$. The displacement $u_j$ of the particles $j=1, \cdots, N$ is then a function of $\xi$,
\begin{equation}
u_j(\xi) = \overline x_j(\xi)-\overline x_j(0),
\end{equation}
where $\overline x_j(\xi)$ and $\overline x_j(0)$ denote the particle position in the minimum energy configuration with and without perturbation, respectively. In the following, we will calculate the displacements $u_j(\xi)$ as a function of the perturbation strength $\xi$.  

Since condition (\ref{equ:minpot}) defines the position of the energy minimum as a function of the perturbation strength $\xi$, its derivative with respect to $\xi$ must vanish,
\begin{equation}
\label{equ:minpotderivation}
\frac{d}{d\xi} \left( \left.\frac{\partial {\mathcal H}(\xi)}{\partial x_j} \right|_{x = \overline x(\xi)} \right) = 0. 
\end{equation}
Application of the chain rule then leads to
\begin{equation}
\label{equ:minpotderivationwithrespecttoa}
\sum_j\left. 
\left(\frac{\partial^2 {\mathcal H}}{\partial x_j \partial x_i} \right|_{x=\overline x(\xi)}\right) 
\frac{\partial \overline x_j(\xi)}{\partial \xi} + 
\left. \frac{\partial^2 {\mathcal H}}{\partial \xi \partial x_i} \right|_{x=\overline x(\xi)} = 0.
\end{equation}
This condition must hold for all $i$. Defining 
\begin{eqnarray}
\label{equ:eigenvaluedefines}
z_j &\equiv& \frac{\partial \overline x_j(\xi)}{\partial \xi},\\
{\mathcal H}_{ij} &\equiv& -\left.\frac{\partial^2 {\mathcal H}}{\partial x_j \partial x_i} \right|_{x=\overline x(\xi)}, \\
\end{eqnarray}
and
\begin{eqnarray}
y_i &\equiv&\left. \frac{\partial^2 {\mathcal H}}{\partial \xi \partial x_i}\right|_{x=\overline x(\xi)} ,
\end{eqnarray}
we can rewrite Equ. (\ref{equ:minpotderivationwithrespecttoa}) as
\begin{equation}
\label{equ:matrixnotation}
y_i = \sum_j {\mathcal H}_{ij}z_j.
\end{equation}
Inversion of the matrix ${\mathcal H}_{ij}$ then yields the vector $z$,
\begin{equation}
\label{equ:matrixnotationinverse}
z_i = \sum_j {\mathcal H}^{-1}_{ij}y_j.
\end{equation}
Once $z_j = \partial \overline x_j(\xi)/\partial \xi$ is known, $\overline x_j(\xi)$ can be obtained by integration.

For the bead-spring model considered here, the first and second derivatives of the potential energy with respect to the particle coordinates are given by
\begin{equation}
\label{equ:firstderivation}
\frac{\partial {\mathcal H}}{\partial x_i} = (2k_1 + k_2)x_i -k_1 x_{i+1} - k_1 x_{i-1} - k_2 b i,
\end{equation}
\begin{equation}
\label{equ:secondderivation}
\frac{\partial {\mathcal H}}{\partial x_i \partial x_j} = \left\{ 
\begin{array}{l l}
  2k_1+k_2 & \quad \mbox{if $i=j$},\\
  -k_1 & \quad \mbox{if $j=i+1$ or $j=i-1$},\\ 
  0 & \quad \mbox{else},\\ 
\end{array} \right. 
\end{equation}
and
\begin{equation}
\label{equ:secondderivationwrta}
\frac{\partial {\mathcal H}}{\partial x_i \partial \xi} = \left\{ 
\begin{array}{l l}
  -k_1 & \quad \mbox{if $i=1$},\\
  0 & \quad \mbox{if $i>0$}.\\ 
\end{array} \right. 
\end{equation}
To solve Equ. (\ref{equ:matrixnotationinverse}) we have to invert the symmetric tridiagonal matrix $H_{ij}$. For this particular matrix, the inverse matrix is known analytically \cite{INVERSEMATRIX},
\begin{eqnarray}
\label{equ:inversematrix}
{\mathcal H}^{-1}_{ij} &=& -\frac{1}{2 k_1} \left\{ \frac{\cosh[(N + 1 - |j-i|)\alpha]}{\sinh(\alpha) \sinh[(N+1)\alpha]} \right. \nonumber \\
& & \hspace{1.0cm }\left.-\frac{\cosh[(N + 1 - i - j)\alpha]}{\sinh(\alpha) \sinh[(N+1)\alpha]} \right\} , 
\end{eqnarray}
where
\begin{equation}
\label{equ:alpha}
\alpha = \cosh^{-1}\left(1 + \frac{k_2}{2k_1} \right)
\end{equation}
Since for our model $y=\{-k_1, 0, 0, \cdots, 0\}$, we obtain
\begin{equation}
\label{equ:inversematrixpartialsolution}
\frac{\partial \overline x_i}{\partial \xi} = \sum_j H^{-1}_{ij} y_j = -H^{-1}_{i0} k_1.
\end{equation}
and hence
\begin{equation}
\frac{\partial \overline x_i}{\partial \xi} = \frac{\cosh[(N + 2 - i)\alpha]-\cosh[(N - i)\alpha]}{2\sinh(\alpha) \sinh[(N+1)\alpha]}.
\end{equation}
For large $N$, this equation simplifies to
\begin{equation}
\frac{\partial \overline x_i}{\partial \xi} = \exp(-i \alpha).
\end{equation}
Integration with respect to $\xi$ then yields
\begin{equation}
\overline x_i(\xi) = \xi \exp(-i \alpha) + C_i,
\end{equation}
where the integration constant is given by $C_i = \overline x_i(0)$. Thus, the displacement of particle $j$ is proportional to the displacement of particle $0$ and decays exponentially with the distance from the origin, 
\begin{equation}
u_i = u_0 \exp(-i \alpha),
\end{equation}
with a decay constant $\alpha$ that depends on the force constants $k_1$ and $k_2$ only.
\end{appendix}

\end{document}